\documentclass[prd,nofootinbib,showpacs,superscriptaddress,11pt]{revtex4}
\usepackage{graphics,epsfig,epsf,psfrag}
\usepackage{amsmath,amssymb,mathrsfs,latexsym}
\usepackage{slashed}
\usepackage{bbm}
\usepackage[usenames]{color}
\usepackage{textcomp,wasysym}
\usepackage{epstopdf}
\usepackage{multirow}
\usepackage{amssymb}
\usepackage{bm}

\newcommand{\gev}{~\mathrm{GeV}}
\newcommand{\mev}{~\mathrm{MeV}}

\newcommand{\x}{X(3872)}
\newcommand{\z}{Z_c(3900)}
\newcommand{\zp}{Z_c(4020)}
\newcommand{\y}{Y(4260)}
\newcommand{\jpc}{J^{PC}}

\newcommand{\ub}{\affiliation{Helmholtz-Institut f\"ur Strahlen- und Kernphysik
       and Bethe Center for Theoretical Physics, Universit\"{a}t Bonn, D-53115
       Bonn, Germany}}
\newcommand{\fzj}{\affiliation{
       Institut f\"{u}r Kernphysik and  Institute for Advanced Simulation,
       Forschungszentrum J\"{u}lich, D--52425 J\"{u}lich, Germany}}
\newcommand{\ihep}{\affiliation{Institute of High Energy Physics and Theoretical
       Physics Center for Science Facilities, Chinese Academy of Sciences,
       Beijing 100049, China}}

\begin{document}

\title{Employing spin symmetry to disentangle different models for the $XYZ$ states}

\author{Martin Cleven}
\email{cleven@ihep.ac.cn}
\ihep
\author{Feng-Kun Guo}
\email{fkguo@hiskp.uni-bonn.de}
\ub
\author{Christoph Hanhart}
\email{c.hanhart@fz-juelich.de}
\fzj
\author{Qian Wang}
\email{q.wang@fz-juelich.de}
\fzj
\author{Qiang Zhao}
\email{zhaoq@ihep.ac.cn}
\ihep

\begin{abstract}
 In order to test different models proposed for some states discovered recently in the
 charmonium mass range that do not fit into the pattern predicted by the conventional
 quark odel,
  we derive predictions for the spectrum within the
 hadro-charmonium picture, the tetraquark picture as well as the hadronic
 molecular approach.
 We exploit heavy quark spin symmetry for the hadro-charmonium and
 hadronic molecule scenarios.
 The patterns that emerge from the different
 models turn out to be quite distinct. For example, only within the
 hadro-charmonium picture a pseudoscalar state emerges that is lighter than
 the $Y(4260)$. Possible discovery channels of these additional states
 are discussed.
\end{abstract}

\date{\today}


\maketitle


\section{Introduction}

In the past decade a number of states in the charmonium and bottomonium sectors
have been discovered with properties in conflict with the quark models for
mesons which were quite successful in describing the low-lying heavy quarkonium
states as $Q\bar Q$ $(Q=c,b)$ states, such as the classical Godfrey--Isgur quark
model~\cite{Godfrey:1985xj} and Cornell potential model~\cite{Eichten:1974af,Eichten:1978tg,Eichten:1979ms}.
The newly observed structures in the heavy quarkonium mass region include,
among many others, the $\x$~\cite{Choi:2003ue},
$\y$~\cite{Aubert:2005rm} and the charged states
$\z$~\cite{Ablikim:2013mio,Liu:2013dau,Xiao:2013iha},
$\zp$~\cite{Ablikim:2013xfr,Ablikim:2013wzq,Ablikim:2013emm,Ablikim:2014dxl} and
$Z_c(4430)$~\cite{Chilikin:2008,Chilikin:2013tch,Aaij:2014jqa} in the charmonium
sector and the $Z_b(10610)$, $Z_b(10650)$\cite{Belle:2011aa}
in the bottomonium sector (see, e.g. the mini review on heavy quarkonium spectroscopy in the Review
of Particle Physics by the PDG~\cite{Agashe:2014kda}).
Because of the failure in describing these structures simply as $Q\bar Q$
mesons, they are considered as candidates of exotic hadrons\footnote{In fact, in
the seminal quark model paper~\cite{GellMann:1964nj}, Gell-Mann already
mentioned the possibility of multiquark states.}. Various models were proposed
to explain these structures, see e.g. Ref.~\cite{Brambilla:2010cs}. However, no
consensus for almost any of these states has been achieved. It is thus of utmost
importance to scrutinize these models, and manifest their distinct predictions.
Because the near-threshold narrow structures in the continuum channel
of the open-flavor meson pair cannot be explained by just a threshold
cusp~\cite{Guo:2014iya}, we will assume that all the states to be discussed
correspond to physical states. The dynamical structure of these states are being
investigated in several different models.
The purpose of this work
 is to present distinct predictions of several commonly discussed models.
The scenarios to be considered include the tetraquark model, the
hadro-charmonium model, and hadronic molecules. A brief description of each
model including a list of relevant references and a short description of the
underlying assumptions will be presented in the corresponding sections.

One has to keep in mind that different components mix with each other whenever
such a mixing is not forbidden. Thus, when we present predictions for a certain scenario for
a given state, we always mean that we are looking for the consequences that arise if
that scenario is assumed to be dominant.

The key feature we are going to exploit in this work is
heavy quark spin symmetry (HQSS). It is an approximate symmetry and becomes exact in the limit of
infinitely heavy quarks. It arises because the spin-dependent quark-gluon
coupling in quantum chromodynamics (QCD) is proportional to the magnetic moment of the
heavy quark, and vanishes in the heavy quark limit. As a result, it leads to
spin multiplets. Within a multiplet, the masses are degenerate in the heavy
quark limit, and the mass splittings depend on the dynamics of the model for
finite quark masses. As will be discussed, the multiplet structure differs in
different scenarios, and thus  provides invaluable information. We will
compare the spectroscopy predicted by each of these three scenarios
in the following sections. We close with a summary.

\section{Hadro-quarkonium}\label{sec:Hadro-quarkonium}

\subsection{Assumptions}

A hadro-quarkonium is a system with a compact heavy quarkonium embedded inside a cloud of
light hadronic matter~\cite{Voloshin:2007dx,Dubynskiy:2008mq}. This scenario was motivated by the fact that
several charmonium-like states were only observed in final states of a specific
charmonium and light hadrons. Examples are the $Y(4260)$
discovered in $J/\psi\pi\pi$~\cite{Aubert:2005rm}, the $Z_c(4430)$
discovered in $\psi'\pi$~\cite{Mizuk:2009da}, the $Y(4360)$ and
$Y(4660)$ observed in $\psi'\pi\pi$~\cite{Wang:2007ea,Aubert:2007zz}.
The recent BESIII observation of similar cross sections for $J/\psi\pi^+\pi^-$
and $h_c\pi^+\pi^-$ at 4.26~GeV and 4.36~GeV in $e^+e^-$
collisions~\cite{Ablikim:2013mio,Ablikim:2013wzq} stimulated Li and Voloshin to generalize the hadro-charmonium
model to include HQSS breaking and describe the $\y$
and $Y(4360)$ as a mixture of two hadro-charmonia~\cite{Li:2013ssa}:
\begin{eqnarray}
  Y(4260) = \cos\theta\,\psi_3 - \sin\theta \, \psi_1\, ,
 \qquad  Y(4360)  = \sin\theta \,\psi_3 + \cos\theta
 \,\psi_1\, ,
 \label{eq:Ymixing}
\end{eqnarray}
where $\psi_1\sim (1^{+-})_{c\bar c}\otimes
(0^{-+})_{q\bar q}$ and  $\psi_3\sim (1^{--})_{c\bar c}\otimes (0^{++})_{q\bar
q}$ are the wave functions of the $J^{PC}=1^{--}$ hadro-charmonia with a
$1^{+-}$ and $1^{--}$ $c\bar c$ core charmonium, respectively. It was argued in
Ref.~\cite{Li:2013ssa} that $\psi_3$ contains predominantly a $\psi'$ (rather than the ground
state $J/\psi$) and that $\psi_1$ contains predominately an $h_c$. The decays into the
$J/\psi \pi^+\pi^-$ then occur through de-exciting $\psi'$ to $J/\psi$ in the
light hadronic matter. The strength is controlled by the so-called
chromo-polarizibility $\alpha_{\psi'J/\psi}$ (see
Ref.~\cite{Sibirtsev:2005ex}).

If we assume that the $J/\psi\pi^+\pi^-$ events seen in $e^+e^-$ collisions  at energies around 4.26 and
4.36~GeV are mainly from the decays of  $\y$ and $Y(4360)$, the BESIII data
imply that $\Gamma(\y\to J/\psi\pi\pi)$ is similar to $\Gamma(Y(4360)\to
J/\psi\pi\pi)$. To achieve this pattern a mixing angle as large as $\theta\approx40^\circ$ is needed~\cite{Li:2013ssa}.
Such a large angle translates into a small mass difference between the $\psi_1$ and $\psi_3$
hadro-charmonia since
\begin{equation}
 \tan (2\theta) = \frac{2 m_{13}}{m_{\psi_1} - m_{\psi_3} }\, ,
 \label{eq:tantheta}
\end{equation}
which can be obtained from 
\begin{equation}
 \left( \begin{array}{cc} m_{Y(4260)} & 0 \\ 0 & m_{Y(4360)}    \end{array}
 \right) =
 \left( \begin{array}{cc} \cos\theta & -\sin\theta \\ \sin\theta  &
 \cos\theta   \end{array} \right) \left( \begin{array}{cc} m_{\psi_1} & m_{13} \\ m_{13} & m_{\psi_3}   \end{array} \right)
 \left( \begin{array}{cc} \cos\theta & \sin\theta \\ -\sin\theta  &
 \cos\theta   \end{array} \right).
\end{equation}
A mixing angle of around $40^\circ$ leads to $m_{\psi_3}\approx4.30$~GeV,
$m_{\psi_1}\approx4.32$~GeV and a mixing amplitude
$m_{13}\approx50$~MeV.

\subsection{Consequences}

In addition to the interference patterns in the line shapes discussed in
Ref.~\cite{Li:2013ssa}, what  else does the proposal of the $Y(4260)$ and $Y(4360)$ as
mixed hadro-charmonia imply? As mentioned in the Introduction, HQSS is quite
useful in this respect. The binding force between the charmonium core and the surrounding light
hadronic matter is due to the exchange of soft gluons.
Because both the charmonium and light hadronic matter are color singlets, at
least two gluons need to be exchanged. The leading order (LO) interaction is due
to exchanging two chromo-electric gluons~\cite{Voloshin:2004un}.
It is important to notice that the heavy (anti-)quark spin decouples from such an
interaction. Therefore, the LO interaction between the light hadronic matter
with a particular heavy quarkonium, [$Q\bar Q$], should be the same as
that for the spin partner of that $[Q\bar Q]$ state. This means that a
hadro-quarkonium should have spin partner(s) just as its core heavy quarkonium does.

Exchanging one chromo-electric and one chromo-magnetic gluon provides a P and
CP odd force, and thus induces the mixing between two hadro-quarkonia with core
heavy quarkonia of opposite P and CP. The mixing between the $\psi_3$ and
$\psi_1$ states, which contain the $\psi'$ ($\text{P}=-,\text{CP}=+$) and the
$h_c$ ($\text{P}=+,\text{CP}=-$), respectively, is such an example.
Exchanging two chromo-magnetic gluons provides one source for splitting the
masses for one spin multiplet of hadro-quarkonia. It is suppressed by
$\mathcal{O}(\Lambda_\text{QCD}^2/m_Q^2)$ in comparison with the LO interaction,
and gives a tiny correction ($\sim 4\%$ for hadro-charmonium and $\sim 1\%$
for hadro-bottomonium) to the potential energy and thus to the mass of the
hadro-quarkonium. The mass splitting between the spin partners within the
same multiplet of hadro-quarkonia is therefore given approximately by the mass
splitting between the core heavy quarkonia with the next-to-leading
spin symmetry violation controlled by mixing analogous to the one
discussed above for the $Y(4260)$ and the $Y(4360)$.
In fact, in Ref.~\cite{Guo:2009id}, analogous to the question of interest, HQSS
has been used to predict that the $Y(4660)$ as a $\psi'f_0(980)$
bound state~\cite{Guo:2008zg} which may be regarded as a specific example of
hadro-charmonium has a spin partner: an $\eta_c'f_0(980)$ bound state, with mass
of around 4616~MeV.

From the discussion above it follows that the $\psi_3$
state has a spin partner characterized by the same light quark
cloud with the $\psi'$ in the core replaced by
the $\eta_c'$. We may call this state  $\eta_{c}[\eta_c']$. It has
$J^{PC}=0^{-+}$ and a mass of around
\begin{equation}
  m_{\psi_3} - (m_{\psi'} - m_{\eta_c'}) \approx
  4.25~\text{GeV}.
\end{equation}
Similarly, replacing the $h_c$ in the $\psi_1$ state by any of the $\chi_{cJ}$ states leads to
three spin partners of  $\psi_1$. The
quantum
numbers of these states composed of
$(J^{++})_{c\bar c}\otimes(0^{-+})_{q\bar q}$ are
$J^{PC}=J^{-+}$.
\begin{table}[tb]
\begin{tabular}{|l c c c|}
    \hline\hline
    Composition \phantom{bla}&  Label & $J^{PC}$~~ 	  & Mass (GeV) \\\hline
    ~~$\psi^\prime\otimes(0^{++})_{q\bar q}$ & $\psi_3$ & $1^{--}$   & 4.30
  \\
    ~~$\eta_c^\prime\otimes(0^{++})_{q\bar q}$ & $\eta_c[\eta_c']$ & $0^{-+}$ &
4.25	 \\\hline
   ~$h_c\otimes(0^{-+})_{q\bar q}$ & $\psi_1$ &  	$1^{--}$  & 4.32
\\
   $\chi_{c0}\otimes(0^{-+})_{q\bar q}$ & $\eta_c[\chi_{c0} ]$  &  $0^{-+}$  &
4.21    \\
   $\chi_{c1}\otimes(0^{-+})_{q\bar q}$ & $\eta_{c1}[\chi_{c1} ]$  &  	$1^{-+}$  & 4.31
  \\
    $\chi_{c2}\otimes(0^{-+})_{q\bar q}$ & $\eta_{c2}[\chi_{c2} ]$  &   $2^{-+}$	  & 4.35
   \\\hline\hline
\end{tabular}
\caption{\label{tab:hadrocharmonium} Masses and quantum numbers of hadro-charmonia in the spin multiplets
of the $\psi_1 \sim h_c\otimes(0^{-+})_{q\bar q}$ and
$\psi_3\sim \psi^\prime\otimes(0^{++})_{q\bar q}$ states.  }
\end{table}
 \begin{figure}[tb]
\centering
\vspace{0cm}
\includegraphics[width=0.5\linewidth]{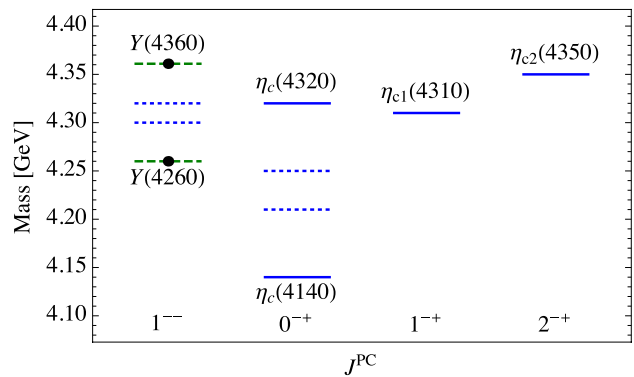}
\vspace{0cm}
\caption{The spectroscopy of the spin partners of $Y(4260)$ and $Y(4360)$
based on the assumption that they are mixed states of two $1^{--}$ hadro-charmonia as proposed in
Ref.~\cite{Li:2013ssa}. The dotted lines
in the vector and pseudoscalar sectors represent the masses of the unmixed states.
The masses of $Y(4360)$ and $Y(4260)$ were used as input for the analysis and are
shown as dashed lines. }
\label{fig:hadrocharmonium}
\end{figure}
Their masses are listed in Table~\ref{tab:hadrocharmonium} and
shown in Fig.~\ref{fig:hadrocharmonium} for illustration.
We notice that there are two states with $\jpc=0^{-+}$ in
analogy to $1^{--}$. As in the vector channel the small mass difference of the pure
spin states of about 40~MeV introduces a sizable mixing.
The relation of the mixing amplitude in this case can be related to that in the
vector case by constructing an CP-odd operator with HQSS breaking
\begin{equation}
  \mathcal{O}_\text{mixing} = \frac14\langle
  \vec{\chi}^{\,\dag}\cdot\vec{\sigma}J' \rangle + \text{h.c.}
  = \vec{h}_c^\dag \cdot \vec{\psi}' + \sqrt{3}\,\chi_{c0}^\dag\eta_c' +
  \text{h.c.}  \,,
\end{equation}
where $\vec{\sigma}$ are the Pauli matrices and $\langle...\rangle$ takes trace
in the spinor space. The fields $J'$ and $\vec{\chi}$ annihilates the $S$-wave
and $P$-wave charmonium states, respectively, and are given explicitly
by~\cite{Fleming:2008yn,Guo:2010ak}
\begin{equation}
J' = \vec{\psi'}\cdot\vec{\sigma}+\eta_c'\,, \qquad
\chi^i = \sigma^j
\left(-\chi_{c2}^{ij}-\frac{1}{\sqrt{2}}\epsilon^{ijk}\chi_{c1}^k +
\frac{1}{\sqrt{3}}\delta^{ij}\chi_{c0} \right) + h_c^i \, .
\end{equation}
Thus the mixing amplitude is $\sqrt{3} m_{13}$,\footnote{We thank M. Voloshin
for pointing this out.} up to corrections of order
$(\Lambda_\text{QCD}/m_c)^2$, with $m_{13}\approx50$~MeV. We can calculate the
eigenvalues of the mass matrix~\footnote{Because these states have pseudoscalar
quantum numbers, we name them as $\eta_c(\text{mass})$.}
\begin{equation}
  m_{\eta_c(4320)} \approx 4.32\gev\, ,~\text{and}~m_{\eta_c(4140)} \approx
4.14\gev\, ,
\end{equation}
and the mixing angle is $\theta_0\approx-38^\circ$ with $\theta_0$ defined via
\begin{eqnarray}
  \eta_c(4320) = \cos\theta_0\,\eta_c[\eta_c'] - \sin\theta_0 \,
\eta_c[\chi_{c0}]\, ,
 \qquad  \eta_c(4140) = \sin\theta_0\,\eta_c[\eta_c'] + \cos\theta_0 \,
\eta_c[\chi_{c0}]\, .
\label{eq:etamixing}
\end{eqnarray}

Please observe that among the states predicted there is a state with exotic quantum
numbers $1^{-+}$: it appears to be a quite robust consequence of the
hadro-charmonium scenario proposed in Ref.~\cite{Li:2013ssa} that an $\eta_{c1}$ state exists with
a mass between those of $Y(4360)$ and $Y(4260)$.

Since all the considered hadro-charmonium states are above the corresponding
thresholds for the decay into the core charmonium and two pions, we expect that
they decay easily through dissociating the light hadronic matter into two
pions, in complete analogy to $Y(4260)$ and $Y(4360)$ that were observed in final
states consisting of a charmonium and two pions. This mechanism will introduce a width of
$\sim100$~MeV, in the ball park of the widths of the $Y(4260)$ and the $Y(4360)$,
for each of them. Considering that both the widths of the $\eta_c'$ and
the $\chi_{c0}$ are about 10~MeV~\cite{Agashe:2014kda}, much larger than those of
the $\psi'$ and the $h_c$, the predicted states $\eta_c(4140)$ and $\eta_c(4320)$
can also decay via the decays of the $\eta_c'$ and $\chi_{c0}$. Hadro-quarkonia
can also decay into open flavor heavy meson and antiheavy meson pairs~\cite{Dubynskiy:2008di}, but it is natural to expect the first mechanism
to be dominant.

In the above, we have argued that if the $Y(4260)$ and  the $Y(4360)$ are mixed
hadro-charmonium states, then it is very likely that they have spin partners
as shown in Fig.~\ref{fig:hadrocharmonium}. Thus searching for these partners
can provide valuable information on the nature of the $Y(4260)$ and $Y(4360)$.
In which processes should they be searched for? As for the two pseudoscalar
states $\eta_c(4140)$ and $\eta_c(4320)$, because of the probably sizeable
mixing, both of them decay into the $\eta_c^{(\prime)}\pi\pi$ and
$\chi_{c0}\pi\pi$. Being pseudoscalars, they cannot be produced directly in
$e^+e^-$ collisions. One way of searching for them is to measure the
$\eta_c^{(\prime)}\pi^+\pi^-$ invariant mass distribution for the decays
$B^\pm\to K^\pm \eta_c^{(\prime)}\pi^+\pi^-$ as suggested in
Ref.~\cite{Guo:2009id} for searching for the spin partner of the $Y(4660)$.
Another possible way of searching for the $\eta_c(4140)$ and $\eta_c(4320)$ is
to study the radiative decays of the $Y(4260)$ and $Y(4360)$. Because the
branching fraction for $\psi'\to\gamma\chi_{c0}$ is
10\%~\cite{Agashe:2014kda}, two orders of magnitude larger than that for
$\psi'\to\gamma\eta_c'$, both the $\eta_c(4140)$ and $\eta_c(4320)$ states can
be produced through the decays of the $\psi_3$ components of the $Y(4260)$ and
$Y(4360)$ into their $\eta_c[\chi_{c0}]$ components. Hence, one may
search for these two states in the process $e^+e^-\to\gamma \chi_{c0}\eta$ at the
center-of-mass energies around the masses of the $Y(4260)$ and the $Y(4360)$.

The exotic state $\eta_{c1}(4310)$ and the state $\eta_{c2}(4350)$
 can be searched for in analogous processes
in the decays of $Y(4360)$ with the $\chi_{c0}$ in the final
state replaced by the $\chi_{c1}$ and $\chi_{c2}$, respectively. This kind of
measurements may be performed at BESIII and a future high-luminosity super
tau-charm factory.

If there are hadro-charmonium states, it is natural that the analogous
hadro-bottomonium states should exist as well~\cite{Dubynskiy:2008mq}. The
interaction strength between the heavy quarkonium and the light hadronic matter
is dictated by the chromo-polarizibility which is a matrix element for the
propagation of a color-octet $Q\bar Q$ pair~\cite{Voloshin:2004un},
and thus depends on the wave functions of the heavy quarkonia involved.
Numerically it was found that the off-diagonal chromo-polarizability for the
transition $\Upsilon'\to \Upsilon\pi\pi$ is a factor of 3 smaller than that for
the charmonium analogue $\psi'\to
J/\psi\pi\pi$~\cite{Sibirtsev:2005ex}.\footnote{For both transitions, the dipion
system is dominantly $S$-wave, and the available phase space for either of them
is larger than 560~MeV for the $\pi\pi$ invariant mass. In this range, one
expects that the $f_0(500)$, thus the $\pi\pi$ final state interaction (FSI),
plays an important role. The FSI was neglected in Ref.~\cite{Sibirtsev:2005ex}.
However, it was found in Ref.~\cite{Guo:2006ya} that the $\pi\pi$ FSI reduces
the values of chromo-polarizability for the $c\bar c$ and $b\bar b$ by a similar
factor. One should also keep in mind that the charged bottomonium-like
$Z_b^\pm(10610)$ and $Z_b^\pm(10650)$ states couple to both the
$\Upsilon\pi^\pm$ and $\Upsilon'\pi^\pm$, hence contribute to the dipion
transition between $\Upsilon'$ and $\Upsilon$. Such a contribution has not been taken into account in the analysis of the chromo-polarizability so
far.} Therefore, there is no flavor symmetry connecting the hadro-bottomnium to the hadro-charmonium.  However, because the mixing is induced by exchanging one chromo-electric and one chromo-magnetic gluon, one naively expects that the mixing amplitude for hadro-bottomonium states is much smaller than that for the hadro-charmonium states, and roughly scales down by a factor of $m_c/m_b$. If the unmixed states do not
accidentally have a tiny mass difference, one should be able to neglect the
mixing between hadro-bottomonium states. This means that the $1^{--}$
hadro-bottomonium states (possibly with a core $b\bar b$ of higher
excitation~\cite{Dubynskiy:2008mq}), should decay cleanly either into
 $\Upsilon(nS)\pi\pi$ or into $h_b(mP)\pi\pi$.

\section{Tetraquark}\label{sec:tetraquark}

\subsection{Assumptions}

Tetraquarks are four-quark states constructed in analogy to the regular
quark model.
In particular, the quarks are held together by effective gluon exchanges. Thus a necessary
feature of each tetraquark model is that each isoscalar state is accompanied by
(nearly) degenerate
isovector states analogous to the $\rho$--$\omega$ degeneracy in the light
meson sector.

Various variants for tetraquark models can be found in the literature. As one representative
of this class of models we here discuss in detail 
 only the implications of the most
recently proposed interaction by Maiani et al.~\cite{Maiani:2014aja} (see also the review
article Ref.~\cite{Esposito:2014rxa}). Note that the interaction originally proposed
for the hidden charm states by the same group~\cite{Maiani:2004vq} was shown to be inconsistent
with the most recent discussions~\cite{Maiani:2014aja,Zhao:2014qva}.
In this model
tetraquarks are understood as such compact diquark--anti-diquark bound systems
that the
spin-spin interactions within the tetraquark are dominated by those within the diquarks.

In this model, the mass of a tetraquark is given by~\cite{Maiani:2014aja}
\begin{eqnarray}
\label{eq:1}
M=M_{00}+B_c\frac{\bm{L}^2}{2}-2a\bm{L}\cdot
\bm{S}+2\kappa_{cq} \left[(\bm{s}_q\cdot \bm{s}_c)+(\bm{s}_{\bar q}\cdot
\bm{s}_{\bar c}) \right] ,
\end{eqnarray}
where $\bm{s}_f (f = q,c,\bar q,\bar c)$ are the spins of (anti-)quarks,
$\bm{S}$ is the total spin, $\bm L$ is the orbital angular momentum between the
diquark and anti-diquark. The (anti-)quarks within the (anti-)diquarks are assumed to be in an $S$--wave.
The parameters $M_{00}$, $B_c$, $a$ and $\kappa_{cq}$ are
 to be fixed from experiment.
Denoting the spin of the diquark and anti-diquark as
$\bm{s} = \bm{s}_q + \bm{s}_c $ and
$\bar{\bm{s}}=\bm{s}_{\bar{q}}+\bm{s}_{\bar{c}}$, respectively,
the Hamiltonian can be evaluated for a given tetraquark state of total angular momentum $J$,
 denoted by
$|s,\bar{s};S,L\rangle_J$,
\begin{equation}
\label{eq:2}
M=M_{00}+B_c\frac{L(L+1)}{2}+a[L(L+1)+S(S+1)-J(J+1)]+
\kappa_{cq} \left[s(s+1)+\bar{s}(\bar{s}+1)-3\right] \ .
\end{equation}
For $J=1$ the expression agree to  Eq.~(38) in Ref.~\cite{Maiani:2014aja}.
Note that the parameters $B_c$, $a$ and $\kappa_{cq}$ are positive values, extracted from the experimental data by Maiani et al. ~\cite{Maiani:2014aja}.
As a result, the mass of the tetraquarks increases with increasing $L$ and $S$,
but decreases for growing $J$, which is a rather unusual feature.

\subsection{Consequences}

A general feature of tetraquark models is that a very rich spectroscopy
emerges. In addition,
 there are always approximately
degenerate isospin singlet and isospin triplet states, analogous to the case of
the $\rho$ and $\omega$ for the traditional $q\bar q$ mesons.

Following Ref.~\cite{Maiani:2014aja} we will discuss the implications of the above model for
$S$-wave and $P$-wave tetraquark states only.
The identification of some of the tetraquark levels with observed states
was presented already in Ref.~\cite{Maiani:2014aja}. Here we extend this investigation
by discussing all possible states with the mentioned quantum numbers.

\begin{figure}
  \centering
  \vspace{0cm}
    \includegraphics[width=1.0\linewidth]{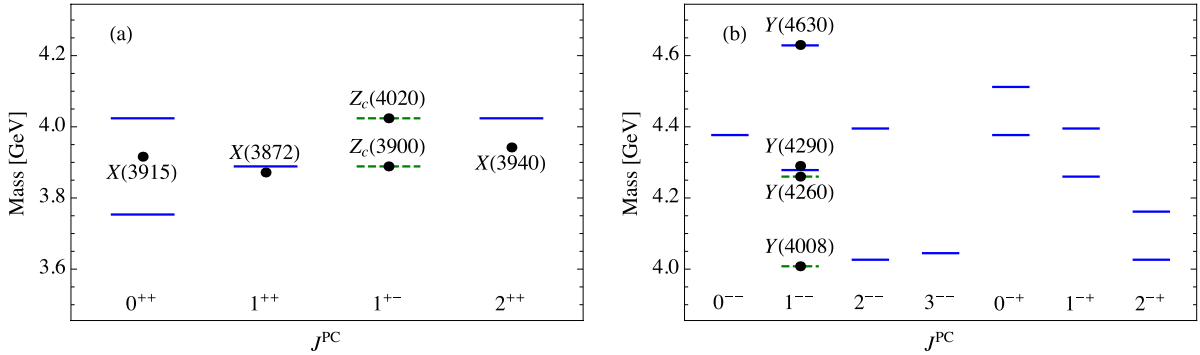}
  \vspace{0cm}
  \caption{The $S$-wave (a) and $P$-wave (b) tetraquark spectroscopy in the
  charmonium sector. The green dashed and blue solid lines are input and
  prediction, respectively. }
  \label{fig:1}
\end{figure}

  For $S$-wave tetraquarks, since $L=0$ and $J=S$ we can use $|s,\bar
  s\rangle_J$ to abbreviate $|s,\bar{s};S,L\rangle_J$.
  Then it follows from Eq.~\eqref{eq:2} that
  there are three sets of tetraquark states, whose masses are
  approximately degenerate within the same set, since $s$ and $\bar s$ are
equal:\footnote{
  The quantum numbers
  are $J^{PC}$, where the $C$-parity is given for the iso-singlet and the neutral
  member of the iso-triplet.}
 \begin{eqnarray}
  \left\{
  \begin{aligned}
    &  0^{++}:~ |0,0\rangle_0 \, ; \\
    &  1^{++}:~ \frac1{\sqrt{2}}\left(|1,0\rangle_1 + |0,1\rangle_1
    \right) , \quad
    & 1^{+-}&:~ \frac1{\sqrt{2}}\left(|1,0\rangle_1 - |0,1\rangle_1
    \right) ; \\
    &  J^{++}:~ |1,1\rangle_0  \, ,\quad
    |1,1\rangle_2 \, ,\quad
    & 1^{+-}&: ~ |1,1\rangle_1 \, .
  \end{aligned}
  \right.
  \end{eqnarray}
  For each of the $J^{PC}$ quantum numbers, as mentioned before, the model predicts  an
  isospin singlet as well as an isospin triplet.\footnote{If we take into account
  the strange and anti-strange quarks, there would be a nonet for each $J^{PC}$.
  } Therefore, there should be in total 24 $S$-wave tetraquark states already even without
considering
  radial excitations that are expected about 400 MeV heavier than the corresponding
  ground states~\cite{Maiani:2014aja}.
  Among the ground states, the authors of
  Ref.~\cite{Maiani:2014aja} identified the $X(3872)$ as one of the neutral
  $1^{++}$ states and the $Z_c(3900)$ and $Z_c(4020)$ as the iso-triplet
  $1^{+-}$ states, $\left(|1,0\rangle_1 - |0,1\rangle_1 \right)/\sqrt{2}$ and
  $|1,1\rangle_1$, respectively. One may also assign $X(3915)$ and $X(3940)$
  as $|1,1\rangle_0$ and $|1,1\rangle_2$,
  respectively~\cite{Maiani:2014aja}, although for each of them there is
  quite a large deviation between the mass of the tetraquark predicted and the actual mass of the observed
  state, cf. Fig.~\ref{fig:1}.
  Therefore, at least 15 of the 24 $S$-wave tetraquarks are waiting for an
observation.

  For $P$-wave tetraquarks, four isospin singlet $1^{--}$ states without radial
  excitation were discussed in Ref.~\cite{Maiani:2014aja}. But there are many
  more states --- a few with exotic quantum numbers (that can not be reached
  by the conventional $\bar qq$ states) like $0^{--}$ and $1^{-+}$
   \begin{eqnarray}
  \left\{
  \begin{aligned}
    & 1^{--}:~ |0,0;0,1\rangle_1 \, ; \\
    & 0^{-+}:~ \frac1{\sqrt{2}} ( |1,0;1,1\rangle_0 +
    |0,1;1,1\rangle_0 )\, , \qquad 0^{--}:~ \frac1{\sqrt{2}} ( |1,0;1,1\rangle_0
    - |0,1;1,1\rangle_0 )\, ; \\
    & 1^{-+}:~ \frac1{\sqrt{2}} ( |1,0;1,1\rangle_1
      + |0,1;1,1\rangle_1 ) \, , \qquad
      1^{--}:~ \frac1{\sqrt{2}} ( |1,0;1,1\rangle_1
      - |0,1;1,1\rangle_1 ) \, ; \\
    & 2^{-+}:~ \frac1{\sqrt{2}} ( |1,0;1,1\rangle_2
      + |0,1;1,1\rangle_2 )\, ,  \qquad
      2^{--}:~ \frac1{\sqrt{2}} ( |1,0;1,1\rangle_2
      - |0,1;1,1\rangle_2 )\, ;  \\
    & 1^{--}:~ |1,1;0,1\rangle_1 \, ; \\
    & 0^{-+}:~ |1,1;1,1\rangle_0\, ; \\
    & 1^{-+}:~ |1,1;1,1\rangle_1 \, ,\qquad  2^{--}:~ |1,1;2,1\rangle_2 \, ;  \\
    & 2^{-+}:~ |1,1;1,1\rangle_2\, ; \\
    & 1^{--}:~ |1,1;2,1\rangle_1\, ; \\
    & 3^{--}:~ |1,1;2,1\rangle_3\, .
  \end{aligned}
  \right.
  \end{eqnarray}According to
  Eq.~\eqref{eq:2} the states in each line are (approximately) degenerate. In
addition, as before,
  all of them appear as an isospin singlet and an
  isospin triplet when we restrict the light quark and anti-quark to be the up
  and down flavors. Thus, the above list amounts to 56 $P$-wave tetraquark
  states even without radial excitations taken into account.

  The assignments of the $1^{--}$ $P$-wave
  tetraquark states to the observed structures in Ref.~\cite{Maiani:2014aja} are as follows: among
  the four $1^{--}$ states, three of them were identified with the $Y(4008)$, $Y(4260)$ and $Y(4630)$,\footnote{It was proposed in the literature
   that $Y(4630)$ observed in the $\Lambda_c\bar \Lambda_c$ and  $Y(4660)$ observed in
  $\psi'\pi^+\pi^-$ correspond to the same state~\cite{Cotugno:2009ys,Guo:2010tk}.} and
  the other one was identified as one of two structures, called $Y(4220)$\footnote{It was proposed in the literature that $Y(4220)$ observed
  $h_c\pi\pi$ and $Y(4260)$ observed in $J/\psi\pi\pi$  correspond to the same state~\cite{Li:2013ssa,Cleven:2013mka}.} and
  $Y(4290)$ observed in $e^+e^-\to h_c\pi^+\pi^-$~\cite{Yuan:2013ffw}. The states $Y(4360)$ and $Y(4660)$ were suggested to be the
  radial excitations of the $Y(4008)$ and $Y(4260)$. Thus, up to the first
  radial excitation, only 6 of 112 (28 if considering only the isospin singlet
  ones) $P$-wave tetraquark states have candidates so far.

  There is one salient feature of Eq.~\eqref{eq:2}: for states with the same
  $s$, $\bar s$, $S$ and $L$, the mass decreases for increasing $J$,
  which appears to be a consequence
  of the negative sign in front of $J(J+1)$.
  For instance, among the states $|1,1;2,1\rangle_J$ with $J^{--}$, the $1^{--}$
  state has the largest mass while the $3^{--}$ one has the smallest.
  Thus the observation of a rather light charmonium with $J=3$ could provide
  strong support for the tetraquark picture of Ref.~\cite{Maiani:2014aja}.

Besides the model discussed in detail above, also other tetraquark models can be found in the literature that
differ in the underlying assumptions. For example, in Refs.~\cite{Hogaasen:2005jv,Buccella:2006fn,Stancu:2009ka,Guo:2011gu,Hogaasen:2013nca}
the states are treated as four-quark systems without any clustering into
diquark--anti-diquark assumed. As a result, the color part of the wave
functions includes both antitriplet and triplet or sextet and anti-sextet configurations for the quark and antiquark pairs,
respectively~\cite{Stancu:2009ka}.
As a result the number of the predicted $S$-wave tetraquarks (Fig.~2 of
Ref.~\cite{Stancu:2009ka}) is twice as large as that of Maiani et al.  Although this picture can explain certain phenomena such as the narrow width of $X(3872)$
due to its tiny $J/\psi+\rho$ and $J/\psi+\omega$ component in the wave
function~\cite{Buccella:2006fn}, $Y(4140)$ as the hidden strange analog
$c\bar c s\bar s$ of $X(3872)$~\cite{Stancu:2009ka}, and $Z_b(10610)$ and $Z_b(10650)$ as $b\bar b
q\bar q$ four-quark systems~\cite{Guo:2011gu}, there is an
even larger number of tetraquarks waiting to be observed within this approach.

\section{Hadronic molecules}\label{sec:molecule}

A hadronic molecule is an extended object that results from non--perturbative
scatterings of two or more hadrons.
The hadronic molecules of interest here are bound states of
a pair of charmed or bottomed mesons, which
are similar to the deuteron as a bound state of the proton and
neutron~\cite{Weinberg:1962hj}. Since the masses of some of the $X, Y, Z$ states
are close to $S$-wave thresholds and couple strongly to the corresponding
continuum states, they are good candidates for hadronic molecules.
For example, the $\x$ is proposed to be a $D\bar{D}^*+c.c.$ molecule~(see
Refs.~\cite{Tornqvist:2004qy,Voloshin:2003nt} and many further studies in the
literature), the $\y$ to be a $D_1\bar{D}+c.c.$
molecule~\cite{Ding:2008gr,Wang:2013cya}, the $Y(4360)$ to be a
$D_1\bar{D}^*+c.c.$ molecule~\cite{Ma:2014zva,Wang:2013kra}, and the two charged states,
$Z_b(10610)^\pm$ and $Z_b(10650)^\pm$,  to be $B\bar{B}^*+c.c.$, $B^*\bar{B}^*$
molecules, respectively \cite{Voloshin:2011qa,Cleven:2011gp}.

Naively,  one might expect that the number of possible molecules is at least as
large as that of the available $S$--wave thresholds.
In addition, since the open-charm and open-bottom mesons carry isospin 1/2  a
pair of them can couple to both isospin 0 and 1.
One might therefore  expect almost degenerate isoscalar and isovector states for
each quantum number similar to the tetraquark scenario.
However, both expectations are not correct. First of all, a shallow bound state
with an unstable constituent can in general not be narrower than that constituent,
but will typically be broader~\cite{Hanhart:2010wh,Filin:2010se}.
In addition, the life time of a broad hadron, whose width is of the order
of or even larger than the inverse of the range of forces, is too short to form
a bound state with another hadron~\cite{Guo:2011dd}.
Thus, only the narrow $D_1(2420)$ with a width of $\sim 25$ MeV can form an
observable hadronic molecule (examples will be discussed below), while the broad
$D_1(2430)$ with a width of $\sim380$~MeV cannot. In this sense it
also appears natural that  the widths of $Y(4260)$ and
$Y(4360)$ are of order $100\mev$.

In addition, the scattering potential in general comprises two contributions: a
long-ranged part mediated by  one-pion exchange and a short-ranged part that is
often parametrized as contact interactions (and that one might be
phenomenologically identified with the exchange of heavier mesons, and could
also come from $s$-channel $\bar qq$ states or more complicated dynamics).
The short-ranged part needs to be fixed from data, as done in
Refs.~\cite{Nieves:2012tt,HidalgoDuque:2012pq,Guo:2013sya} for the
systems of a pair of $S$-wave heavy and anti-heavy mesons using the information of the $X(3872)$ and
$Z_b(10610)$ states as input.
 Since at present there is not enough experimental information available for doing 
this for all channels, in this Section we have to
restrict ourselves to qualitative statements regarding the hadronic molecular
picture.\footnote{For the discussion of the one-pion exchange in effective
field theories for the $X(3872)$, see
Refs.~\cite{Fleming:2007rp,Nieves:2011vw,Valderrama:2012jv,Nieves:2012tt,
Baru:2015nea,Alhakami:2015uea,Braaten:2015tga}.
In addition, there exist various model calculations based on one-meson exchange
potential or SU(4) extension of the light meson interactions.
For the systems of a pair of $S$-wave heavy mesons, see, e.g.,
Refs.~\cite{Tornqvist:1993ng,Zhang:2006ix,Thomas:2008ja,Liu:2008tn,Liu:2008mi,
Molina:2009ct,Lee:2009hy,Sun:2011uh,Ohkoda:2011vj,Ozpineci:2013zas,He:2013nwa,
Aceti:2014kja,Aceti:2014uea,Zhao:2014gqa}. For the systems of an $S$-wave and a
$P$-wave heavy meson, see
Refs.~\cite{Liu:2008xz,Ding:2008mp,Ding:2008gr,Li:2015exa}.
} However, this has already allowed us to highlight some striking differences in the features of this model in comparison with the tetraquarks and hadro-charmonia.

As a guidance for the existence of shallow bound hadronic molecules (here, we
wish to use the phrase ``shallow" in a loose sense meaning states with binding
energies significantly less than 100 MeV), we will study the contribution of the
one-pion exchange and argue that if the one-pion exchange is repulsive for a
given system the appearance of a bound state is unlikely, while
a bound state could exist for an attractive one-pion exchange.\footnote{Note
that in Ref.~\cite{Aceti:2014uea} it is claimed that one-pion exchange does not
contribute to the binding of, e.g., $Z_c(3900)$, since it gets cancelled by the
contribution of the $\eta$ and $\eta'$ exchanges in the $U(3)$ limit. However,
especially in the $D^*\bar D$ system where the exchanged pion is near on-shell
while $\eta$ and $\eta'$ are far off-shell, one should expect sizable violations
of $U(3)$ symmetry.}
 This kind of argument is justified by the very small mass of the pion together
 with the observation that an attractive interaction mediated by a massless
 exchange particle always binds, regardless how weak the interaction
 is~\cite{grossbuch}. We also notice the argument by
 Eriscon and Karl~\cite{Ericson:1993wy} suggesting that two hadrons with
 an attractive one-pion exchange potential should form hadronic molecules if
 their reduced mass is sufficiently large.
 Based on this kind of reasoning it was possible to predict the existence of the $X(3872)$ well before its observation~\cite{Tornqvist:1991ks}.
 There are also examples discussed below
where the one-pion exchange does not contribute and possible molecular states must
then be bound either by coupled channel effects or by short-ranged interactions.
In such a case, the reasoning used in this paper cannot be applied.

As mentioned above when talking about bound systems of mesons only those meson
pairs are relevant where both mesons are sufficiently narrow. Here we focus on
 the ground state open charm meson doublet $D$ and $D^*$ (characterized
by a charm quark and a light anti-quark contribution with $s_\ell^P=\frac12^-$
where $s_\ell$ is the total angular momentum of the light part) and the spin doublet
that contains $D_1(2420)$ and $D_2(2460)$ (characterized by a charm quark and a
light anti-quark contribution with total $s_\ell^P=\frac32^+$) and their
anti-particles as possible constituents. In particular, we discuss the $\frac12
+ \frac 12$ and $\frac12 + \frac 32$ hadronic molecules, where we have used the
total angular momentum of the light quark contribution to characterize the states.

\begin{figure}
\vspace{0cm}
\includegraphics[width=0.8\linewidth]{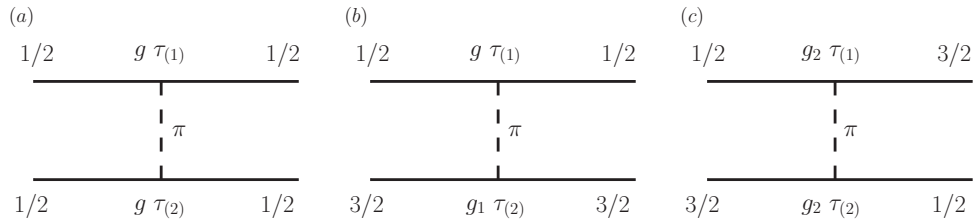}
\vspace{0cm}
\caption{The various one--pion exchange diagrams relevant for the hadronic molecular scenario.
Diagrams (a) and be show the $t$--channel pion exchange for the $1/2+1/2$ and
the $1/2+3/2$ system, respectively, while diagram (c) shows the $u$--channel exchange, relevant for
the $1/2+3/2$ system only.}
\label{fig:4}
\end{figure}

The Feynman diagram of the one-pion exchange is shown in Fig.~\ref{fig:4}.
Because the pions are in the adjoint representation of the isospin $SU(2)$
group, and the non-strange heavy mesons are isospin-1/2 states, each
heavy-meson--pion vertex is accompanied by a factor
$\vec\tau$ which stands for the Pauli matrices operating in the isospin space. The full one-pion exchange
contribution between two heavy mesons therefore comes with a factor $\vec \tau_{(1)}
\cdot \vec \tau_{(2)}$,  where the subindices label the heavy meson to which the
corresponding vertex is attached. One finds for a given pair of isospin-1/2
particles with total isospin $I$ and the third component $I_3$
\begin{equation}
\langle I I_3| \vec \tau_{(1)} \cdot \vec \tau_{(2)} | I I_3\rangle =
2\left[I(I+1)-\frac32\right],
\end{equation}
which is $-3$ for $I=0$ and 1 for $I=1$.
Thus, if for a given set of quantum numbers the one-pion exchange contribution
is attractive in the isoscalar channel, it will be repulsive in the isovector,
and vise versa. Consequently, there is either an isoscalar state \emph{or} an
isovector state with some fixed quantum numbers, but typically not both. This is
in contrast to the predictions for the tetraquark
models discussed in the previous section, since the one-gluon exchange
potential is flavor independent.

In this paper we discuss where to expect $S$-wave molecular states from
non-perturbative interactions of the $\frac12^-$ multiplet $(D,D^*)$ with the
anti-particles $(\bar D, \bar D^*)$ or with the $\frac32^+$ multiplet $(\bar
D_1, \bar D_2)$. Particles and anti-particles need to be combined such that the
states  have a well defined $C$-parity. Bound systems formed from members of two
$\frac32^+$ multiplets are potentially too broad to show a striking signal in
experiment and will therefore not be discussed here. Within this model space the
following quantum numbers can be reached for $\frac12+\frac12$,
\begin{equation}
\begin{aligned}
\displaystyle   0^{++}:& \quad | D^*\bar{D}^*\rangle , \quad | D\bar{D}\rangle\, ; \\
  \displaystyle 1^{+-}:& \quad \frac{1}{\sqrt{2}}| D\bar{D}^*-\bar{D}D^*\rangle
    , \quad | D^*\bar{D}^*\rangle \, ;  \\
    \displaystyle 1^{++}:& \quad
    \frac{1}{\sqrt{2}}| D\bar{D}^*+\bar{D}D^*\rangle \, ; \\
 \displaystyle 2^{++}:& \quad  | D^*\bar{D}^*\rangle\, .
\end{aligned}
\end{equation}
Here and in the following, we use the phase convention for the charge
conjugation so that $\bar D_J = \mathcal{C} D_J \mathcal{C}^{-1}$ with $D_J$
representing any charmed meson and $\mathcal{C}$ the charge conjugation
operator.
Whenever there appears more than one state for given quantum numbers, hadronic
molecular states may appear as a result of coupled-channel dynamics and very limited statements
are possible without a detailed dynamical calculation. Exception to this are the $1^{++}$ and $2^{++}$ states. It is natural to
identify the $1^{++}$ state with the $X(3872)$.
In addition, it turned out that HQSS forces the binding potentials for these two
cases to be equal at LO in the low-energy expansion such that one arrives at the
prediction of a tensor state $X_2$ located close to the $D^*\bar D^*$
threshold~\cite{Nieves:2012tt,Guo:2013sya}.
In Ref.~\cite{Albaladejo:2015dsa} it was shown that within a scheme where the
one-pion exchange is treated perturbatively, this result is stable under
inclusion of the $D\bar D$ inelastic channel in the $D$ wave. As mentioned
above, as a result of the isospin factor of the one-pion exchange, in the
molecular picture one does not expect any isovector states with the quantum
numbers $1^{++}$ and $2^{++}$.

Coupled-channel equations based on LO $S$-wave interactions typically produce as
many states as the channels included in the calculation. Thus, we expect two
states to be present with quantum numbers $1^{+-}$ that strongly couple to
$D\bar D^*$ and $D^*\bar D^*$. Indeed, both are established experimentally with
the isovector states $Z_c(3900)$ and $Z_c(4020)$. Current experimental evidence
locates both states above the thresholds~\footnote{It should be stressed that the
location of the poles is not settled yet: that the peaks in the data are located above the
threshold does not necessarily imply that the poles are above the threshold as well. This
is demonstrated for the $Z_b$--states in Ref.~\cite{Cleven:2011gp}.}. For $S$-wave states this can
be achieved either by momentum dependent interactions~\cite{Hanhart:2014ssa} or
non-trivial coupled-channel dynamics.
Which one of these mechanisms is at work here (if any) requires a detailed
model-building which however is beyond the scope of this work. Again, since isovector
states are observed one should not expect their isoscalar partner states with
$1^{+-}$ quantum numbers.

The situation for $0^{++}$ is more complicated, since also for this system we
are faced with a coupled-channel system.
 In addition, the one-pion exchange is not allowed for the diagonal
$D\bar D$ interaction as a consequence of parity conservation. Neither does HQSS
equalize the LO contact interaction in this channel to that in any other
channel~\cite{Nieves:2012tt}. Thus, at this point we cannot make any statement
about the existence or non-existence of hadronic molecular states in the $0^{++}$
channel.

The number of available channels and quantum numbers for the $\frac12+\frac32$
system is even much larger than the one for $\frac12+\frac12$:
\begin{eqnarray}
\begin{aligned}
  \displaystyle  1^{-+}: &\quad  \frac{1}{\sqrt 2}|D\bar{D}_1  + D_1\bar{D}\rangle\, , \quad
    {\frac{1}{\sqrt 2}}|D^*\bar{D}_1+ D_1\bar{D}^*\rangle\, , \quad
    \frac{1}{\sqrt 2}|D^*\bar{D}_2+ D_2\bar{D}^*\rangle \, ; \\
    \displaystyle1^{--}: &\quad \frac{1}{\sqrt 2}|D\bar{D}_1  - D_1\bar{D}\rangle\, , \quad
   {\frac{1}{\sqrt 2}}|D^*\bar{D}_1- D_1\bar{D}^*\rangle\, , \quad
   \frac{1}{\sqrt 2}|D^*\bar{D}_2- D_2\bar{D}^*\rangle \, ; \\
   \displaystyle 2^{-+}: &\quad \frac{1}{\sqrt 2}|D\bar{D}_  2+ D_2\bar{D}\rangle\, , \quad
    \frac{1}{\sqrt 2}|D^*\bar{D}_1+ D_1\bar{D}^*\rangle\, , \quad
    \frac{1}{\sqrt 2}|D^*\bar{D}_2+ D_2\bar{D}^*\rangle \, ;\\
  \displaystyle  2^{--}: &\quad \frac{1}{\sqrt 2}|D\bar{D}_  2- D_2\bar{D}\rangle\, , \quad
    \frac{1}{\sqrt 2}|D^*\bar{D}_1- D_1\bar{D}^*\rangle\, , \quad
    \frac{1}{\sqrt 2}|D^*\bar{D}_2- D_2\bar{D}^*\rangle \, ;\\
 \displaystyle   0^{-+}: &\quad \frac{1}{\sqrt 2}|D^*\bar{D}_1+ D_1\bar{D}^*\rangle \, ;\\
 \displaystyle   0^{--}: &\quad \frac{1}{\sqrt 2}|D^*\bar{D}_1- D_1\bar{D}^*\rangle \, ;\\
 \displaystyle   3^{-+}: &\quad \frac{1}{\sqrt 2}|D^*\bar{D}_2+ D_2\bar{D}^*\rangle\, ;\\
 \displaystyle   3^{--}: &\quad \frac{1}{\sqrt 2}|D^*\bar{D}_2- D_2\bar{D}^*\rangle\, .
 \end{aligned}
\end{eqnarray}

In this case, one can also check in which channels HQSS predicts the same LO
interaction. For each isospin, 0 or 1, there are four independent interactions
denoted as $\langle
s_{\ell1},s_{\ell2},s_L|\hat V_I|s_{\ell1}',s_{\ell2}',s_L\rangle$ with
$\bm{s}_L=\bm{s}_{\ell1}+\bm{s}_{\ell2}=\bm{s}_{\ell1}'+\bm{s}_{\ell2}'$:
\footnote{We thank J. Nieves for pointing this out.}
\begin{eqnarray}
  \left\langle \frac12,\frac32,1 \left|\hat V_I \right|\frac12,\frac32,1 \right\rangle , \
   \left\langle \frac12,\frac32,1 \left|\hat V_I\right|\frac32,\frac12,1 \right\rangle , \
   \left\langle \frac12,\frac32,2 \left|\hat V_I\right|\frac12,\frac32,2 \right\rangle , \
   \left\langle \frac12,\frac32,2 \left|\hat V_I\right|\frac32,\frac12,2 \right\rangle .
  \label{eq:VHT}
\end{eqnarray}
It turns out that for the diagonal interactions all of the above listed
channels have a different linear combination of the matrix elements given in
Eq.~\eqref{eq:VHT}. Therefore, not much can be derived from HQSS without further
input in this case.

\begin{figure}
  \centering
  \vspace{0cm}
    \includegraphics[width=0.6\linewidth]{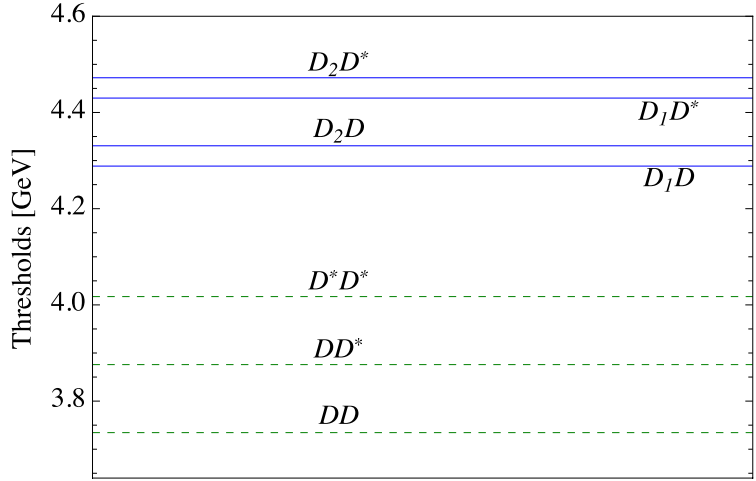}
  \vspace{0cm}
  \caption{The two--body thresholds in the charmonium mass range potentially relevant for the
  formation of hadronic molecules. }
  \label{fig:molecules}
\end{figure}

In addition, in case of the $\frac12+\frac32$ system two kinds of one-pion exchange
contributions are possible in the channels with odd parity. They correspond to
the $t$-channel and $u$-channel one-pion exchange, cf. Fig.~\ref{fig:4}.
Denoting the coupling constants for the coupling of a pion with two $\frac12$ states ($P$-wave), with two $\frac32$ states ($P$-wave) and
connecting a $\frac12$ state and a $\frac32$ state ($D$-wave)  by $g$, $g_1$ and $g_2$, respectively, the potential
for the $t$-channel exchange is proportional to $gg_1$ and the one for the
$u$-channel exchange is proportional to $g_2^2$.
While the magnitude and sign of the coupling constant $g$ are fixed by data and
lattice QCD calculations, nothing is known yet about $g_1$. Therefore, we cannot
identify the channels where the corresponding one-pion exchange potential is
attractive. The situation is different for the part of $u$-channel exchange with
two $D$-wave vertices proportional to $g_2^2$ (although the additional momentum
factors on the vertices might suppress this part of the potential). Looking only at the sign of
this $u$-channel exchange contribution for the uncoupled channels, one finds
attraction for isoscalar states with the exotic quantum numbers $0^{--}$ and
$3^{-+}$, while this part of the interaction is attractive in the isovector
channel with the quantum numbers $0^{-+}$ and $3^{--}$. However, as mentioned
above, since for these systems we cannot even analyze the complete one-pion
exchange contribution no strong conclusion can be drawn regarding the existence
or non-existence of these states.

For the quantum numbers  $1^{-\pm}$, $2^{-\pm}$  three channels couple to each
other. Thus, here even less can be said without dynamical analysis. However,
some general statements are still possible. For instance, since hadronic
molecular states appear in the vicinity of $S$-wave thresholds, the lightest molecular
state in the $\frac12+\frac32$ family is expected to be close to the lowest
threshold, namely $D_1\bar D$ (cf. Fig.~\ref{fig:molecules}), and with the
quantum numbers of either $1^{--}$ or $1^{-+}$. Indeed, in
Refs.~\cite{Ding:2008gr,Wang:2013cya} the $Y(4260)$ was proposed to be
(predominantly) a $D_1\bar D$ bound state with $J^{PC}=1^{--}$, and in
Ref.~\cite{Wang:2014wga} the molecular states with exotic quantum numbers
$J^{PC}=1^{-+}$ were investigated.

For the quantum numbers $0^{-\pm}$, if a corresponding state exists at all, it
should be around the $D^*\bar D_1$ threshold and thus more than 100 MeV above
the mass of the $Y(4260)$. While this mass range is similar to what is predicted
in the tetraquark picture, it is very different to what is predicted in the
hadro-charmonium scenario, where the $0^{-+}$ states are predicted to be the
lightest of their kind.

One striking distinction between the tetraquark picture and the
hadronic molecular scenario becomes visible for the $J=3$ states: while in the
tetraquark model of Ref.~\cite{Maiani:2014aja} those are amongst the
lightest states, in the
hadronic molecular picture, if they exist, they should be close to the $D_2\bar
D^*$ threshold, again more than 100~MeV above the $Y(4260)$, cf.
Fig.~\ref{fig:molecules}.

\section{Summary}\label{sec:summary}

In this work we have investigated the exotic charmonium spectrum in different
scenarios based on HQSS. In particular, we have compared the
spectra arising from three models, i.e. hadro-charmonium, tetraquark and hadronic molecule,
which turn out to provide quite distinct predictions for not yet discovered
``$XYZ$" states.

In the hadro-charmonium scenario we find that if $Y(4260)$ and $Y(4360)$ are mixed hadro-charmonium states with $J^{PC}=1^{--}$
as proposed in Ref.~\cite{Li:2013ssa}
then two  spin partners with $J^{PC}=0^{-+}$ and two more with  $J^{PC}=1^{-+}$ and  $J^{PC}=2^{-+}$, respectively,  should exist as well.
Their possible production and decay modes are discussed to guide the experimental search in the future.

In the tetraquark scenario  the spectrum is much richer than in other scenarios
with approximate degeneracies between isospin singlet and triplet states.
If one assigns the observed $X(3872)$, $Z_c(3900)$ and $Z_c(4020)$ to the
$S$-wave states, some of the states with $J^{PC}=1^{--}$ can also be assigned to
the $P$-wave tetraquark states, i.e. $Y(4008)$, $Y(4260)$ and $Y(4630)$, while
the possible existence of two $1^{--}$ states, $Y(4220)$ and $Y(4290)$ in
$e^+e^-\to h_c \pi^+\pi^-$ will accommodate all the $P$-wave $1^{--}$ tetraquark
ground states. However, it should be noted that so far only a small number of
predicted tetraquark states can be assigned to existing observations ---
at least 67 among 80 of the ground states are left to be discovered, if compact
tetraquarks provide the dominant components of the new $XYZ$ states.

In the scenario of hadronic molecule very little can be said without detailed
modeling of the relevant hadron-hadron potentials.
Under the assumption
 that the one-pion exchange potential  plays a dominant role, one expects bound
 states to appear either in the isoscalar or isovector channel but not in both
 simultaneously. The observation of the isospin singlet $X(3872)$ and the isospin triplet
 $Z_c(3900)$ and $Z_c(4020)$ without evidence of isospin partners matches this
 expectation well. Taking into account that the constituent mesons are to be
 narrow enough in the hadronic molecular system, the allowed number for
 hadronic molecules formed by low-lying narrow charmed meson pairs turns out to
 be a lot smaller than what is predicted within the tetraquark scenario.

 In order to highlight that the spectroscopy of the different scenarios is
 indeed very different we summarize the implications for two quantum numbers:
 Amongst the three models only the hadro-charmonium scenario predicts 2
 $0^{-+}$ states with one of them being even lighter than the $Y(4260)$. In the
 tetraquark picture also two $0^{-+}$ states appear, however, both are predicted
 to be heavier than $Y(4260)$.
 Due to lack of a dynamical model, in the hadronic molecular picture there is no
 prediction yet for a $0^{-+}$ state, however, if it exists it should be located
 near the $D_1\bar D^*$ threshold --- again above the mass of the $Y(4260)$.

Interestingly, we find that the tetraquark model of
Ref.~\cite{Maiani:2014aja} and hadronic molecular scenarios
predict strikingly different patterns for the $J=3$ states. In the former
scenario the $J^{PC}=3^{--}$ state is expected to be among the lightest states,
while in the hadronic molecular scenario the $J=3$ states should have masses
close to the $D_2\bar{D}^*$ threshold if they exist. An observation of the $J=3$
states should provide crucial information about the underlying dynamics.

It is very important to test the different scenarios in future experimental
studies at LHCb, BESIII, Belle-II, PANDA and others.

\section*{Acknowledgement}

We are very grateful to J. Nieves and M. Voloshin for discussions and valuable
comments.
This work is supported, in part, by the NSFC and DFG
through funds provided to the Sino-German CRC 110 ``Symmetries and
the Emergence of Structure in QCD" (NSFC Grant No. 11261130311),
NSFC (Grant Nos. 11425525 and 11165005), and the Ministry of Science and Technology of China
(2015CB856700).
MC is also supported by the Chinese Academy of Sciences President's International Fellowship Initiative grant 2015PM006.


\begin{thebibliography}{99}


\bibitem{Godfrey:1985xj}
  S.~Godfrey and N.~Isgur,
  Phys.\ Rev.\ D {\bf 32}, 189 (1985).

\bibitem{Eichten:1974af}
  E.~Eichten, K.~Gottfried, T.~Kinoshita, J.~B.~Kogut, K.~D.~Lane and T.~M.~Yan,
  Phys.\ Rev.\ Lett.\  {\bf 34}, 369 (1975)
  [Phys.\ Rev.\ Lett.\  {\bf 36}, 1276 (1976)].

\bibitem{Eichten:1978tg}
  E.~Eichten, K.~Gottfried, T.~Kinoshita, K.~D.~Lane and T.~M.~Yan,
  Phys.\ Rev.\ D {\bf 17}, 3090 (1978)
  [Phys.\ Rev.\ D {\bf 21}, 313 (1980)].
  
\bibitem{Eichten:1979ms}
  E.~Eichten, K.~Gottfried, T.~Kinoshita, K.~D.~Lane and T.~M.~Yan,
  Phys.\ Rev.\ D {\bf 21}, 203 (1980).


\bibitem{Choi:2003ue}
  S.~K.~Choi {\it et al.}  [Belle Collaboration],
  Phys.\ Rev.\ Lett.\  {\bf 91}, 262001 (2003)
  [hep-ex/0309032].

\bibitem{Aubert:2005rm}
  B.~Aubert {\it et al.}  [BaBar Collaboration],
  Phys.\ Rev.\ Lett.\  {\bf 95}, 142001 (2005)
  [hep-ex/0506081].

\bibitem{Ablikim:2013mio}
  M.~Ablikim {\it et al.}  [BESIII Collaboration],
  Phys.\ Rev.\ Lett.\  {\bf 110}, 252001 (2013)
  [arXiv:1303.5949 [hep-ex]].

\bibitem{Liu:2013dau}
  Z.~Q.~Liu {\it et al.}  [Belle Collaboration],
  Phys.\ Rev.\ Lett.\  {\bf 110}, 252002 (2013)
  [arXiv:1304.0121 [hep-ex]].

\bibitem{Xiao:2013iha}
  T.~Xiao, S.~Dobbs, A.~Tomaradze and K.~K.~Seth,
  Phys.\ Lett.\ B {\bf 727}, 366 (2013)
  [arXiv:1304.3036 [hep-ex]].

\bibitem{Ablikim:2013xfr}
  M.~Ablikim {\it et al.}  [BESIII Collaboration],
  Phys.\ Rev.\ Lett.\  {\bf 112},  022001 (2014)
  [arXiv:1310.1163 [hep-ex]].

\bibitem{Ablikim:2013wzq}
  M.~Ablikim {\it et al.}  [BESIII Collaboration],
  Phys.\ Rev.\ Lett.\  {\bf 111},  242001 (2013)
  [arXiv:1309.1896 [hep-ex]].

\bibitem{Ablikim:2013emm}
  M.~Ablikim {\it et al.}  [BESIII Collaboration],
  Phys.\ Rev.\ Lett.\  {\bf 112},  132001 (2014)
  [arXiv:1308.2760 [hep-ex]].

\bibitem{Ablikim:2014dxl}
  M.~Ablikim {\it et al.}  [BESIII Collaboration],
  Phys.\ Rev.\ Lett.\  {\bf 113},  212002 (2014)
  [arXiv:1409.6577 [hep-ex]].

\bibitem{Chilikin:2008}
S. K. Choi et al. [Belle Collaboration], Phys. Rev. Lett.
100, 142001 (2008).

\bibitem{Chilikin:2013tch}
K.~Chilikin {\it et al.}  [Belle Collaboration],
  Phys.\ Rev.\ D {\bf 88},  074026 (2013)
  [arXiv:1306.4894 [hep-ex]].

\bibitem{Aaij:2014jqa}
  R.~Aaij {\it et al.}  [LHCb Collaboration],
  Phys.\ Rev.\ Lett.\  {\bf 112},  222002 (2014)
  [arXiv:1404.1903 [hep-ex]].

\bibitem{Belle:2011aa}
  A.~Bondar {\it et al.}  [Belle Collaboration],
  Phys.\ Rev.\ Lett.\  {\bf 108}, 122001 (2012)
  [arXiv:1110.2251 [hep-ex]].

\bibitem{Agashe:2014kda}
  K.~A.~Olive {\it et al.}  [Particle Data Group],
  Chin.\ Phys.\ C {\bf 38}, 090001 (2014).

 \bibitem{GellMann:1964nj}
  M.~Gell-Mann,
  Phys.\ Lett.\  {\bf 8}, 214 (1964).


\bibitem{Brambilla:2010cs}
  N.~Brambilla {\it et al.},
  Eur.\ Phys.\ J.\ C {\bf 71}, 1534 (2011)
  [arXiv:1010.5827 [hep-ph]].

\bibitem{Guo:2014iya}
  F.-K.~Guo, C.~Hanhart, Q.~Wang and Q.~Zhao,
  Phys.\ Rev.\ D {\bf 91}, 051504 (2015)
  [arXiv:1411.5584 [hep-ph]].

\bibitem{Voloshin:2007dx}
  M.~B.~Voloshin,
  Prog.\ Part.\ Nucl.\ Phys.\  {\bf 61}, 455 (2008)
  [arXiv:0711.4556 [hep-ph]].

\bibitem{Dubynskiy:2008mq}
  S.~Dubynskiy and M.~B.~Voloshin,
  Phys.\ Lett.\ B {\bf 666}, 344 (2008)
  [arXiv:0803.2224 [hep-ph]].

\bibitem{Mizuk:2009da} 
  R.~Mizuk {\it et al.}  [Belle Collaboration],
  Phys.\ Rev.\ D {\bf 80}, 031104 (2009)
  [arXiv:0905.2869 [hep-ex]].

\bibitem{Wang:2007ea} 
  X.~L.~Wang {\it et al.}  [Belle Collaboration],
  Phys.\ Rev.\ Lett.\  {\bf 99}, 142002 (2007)
  [arXiv:0707.3699 [hep-ex]].

\bibitem{Aubert:2007zz}  
  B.~Aubert {\it et al.}  [BaBar Collaboration],
  Phys.\ Rev.\ Lett.\  {\bf 98}, 212001 (2007)
  [hep-ex/0610057].

\bibitem{Li:2013ssa}
  X.~Li and M.~B.~Voloshin,
  Mod.\ Phys.\ Lett.\ A {\bf 29},  1450060 (2014)
  [arXiv:1309.1681 [hep-ph]].

\bibitem{Sibirtsev:2005ex}
  A.~Sibirtsev and M.~B.~Voloshin,
  Phys.\ Rev.\ D {\bf 71}, 076005 (2005) [hep-ph/0502068].

 \bibitem{Voloshin:2004un}
  M.~B.~Voloshin,
  Mod.\ Phys.\ Lett.\ A {\bf 19}, 665 (2004)
  [hep-ph/0402011].

\bibitem{Guo:2009id}
  F.-K.~Guo, C.~Hanhart and U.-G.~Mei\ss ner,
  Phys.\ Rev.\ Lett.\  {\bf 102}, 242004 (2009)
  [arXiv:0904.3338 [hep-ph]].

\bibitem{Fleming:2008yn}
  S.~Fleming and T.~Mehen,
  Phys.\ Rev.\ D {\bf 78}, 094019 (2008)
  [arXiv:0807.2674 [hep-ph]].

\bibitem{Guo:2008zg}
  F.-K.~Guo, C.~Hanhart and U.-G.~Mei\ss ner,
  Phys.\ Lett.\ B {\bf 665}, 26 (2008)
  [arXiv:0803.1392 [hep-ph]].

\bibitem{Guo:2010ak}
  F.-K.~Guo, C.~Hanhart, G.~Li, U.-G.~Mei\ss{}ner and Q.~Zhao,
  Phys.\ Rev.\ D {\bf 83}, 034013 (2011)
  [arXiv:1008.3632 [hep-ph]].

\bibitem{Dubynskiy:2008di}
  S.~Dubynskiy, A.~Gorsky and M.~B.~Voloshin,
  Phys.\ Lett.\ B {\bf 671}, 82 (2009)
  [arXiv:0804.2244 [hep-th]].

\bibitem{Guo:2006ya}
  F.-K.~Guo, P.-N.~Shen and H.-C.~Chiang,
  Phys.\ Rev.\ D {\bf 74}, 014011 (2006)
  [hep-ph/0604252].

\bibitem{Maiani:2014aja}
  L.~Maiani, F.~Piccinini, A.~D.~Polosa and V.~Riquer,
  Phys.\ Rev.\ D {\bf 89},  114010 (2014)
  [arXiv:1405.1551 [hep-ph]].


\bibitem{Esposito:2014rxa}
  A.~Esposito, A.~L.~Guerrieri, F.~Piccinini, A.~Pilloni and A.~D.~Polosa,
  Int.\ J.\ Mod.\ Phys.\ A {\bf 30}, 1530002 (2015)
  [arXiv:1411.5997 [hep-ph]].

\bibitem{Maiani:2004vq}
  L.~Maiani, F.~Piccinini, A.~D.~Polosa and V.~Riquer,
  Phys.\ Rev.\ D {\bf 71}, 014028 (2005)
  [hep-ph/0412098].

\bibitem{Zhao:2014qva}
  L.~Zhao, W.~Z.~Deng and S.~L.~Zhu,
  Phys.\ Rev.\ D {\bf 90},  094031 (2014)
  [arXiv:1408.3924 [hep-ph]].

\bibitem{Cotugno:2009ys}
  G.~Cotugno, R.~Faccini, A.~D.~Polosa and C.~Sabelli,
  Phys.\ Rev.\ Lett.\  {\bf 104} (2010) 132005
  [arXiv:0911.2178 [hep-ph]].

\bibitem{Guo:2010tk}
  F.-K.~Guo, J.~Haidenbauer, C.~Hanhart and U.-G.~Mei\ss ner,
  Phys.\ Rev.\ D {\bf 82}, 094008 (2010)
  [arXiv:1005.2055 [hep-ph]].

\bibitem{Cleven:2013mka}
  M.~Cleven, Q.~Wang, F.-K.~Guo, C.~Hanhart, U.-G.~Mei\ss ner and Q.~Zhao,
  Phys.\ Rev.\ D {\bf 90} (2014) 7,  074039
  [arXiv:1310.2190 [hep-ph]].

\bibitem{Yuan:2013ffw}
  C.~Z.~Yuan,
  Chin.\ Phys.\ C {\bf 38}, 043001 (2014)
  [arXiv:1312.6399 [hep-ex]].

\bibitem{Hogaasen:2005jv}
  H.~H{\o}gaasen, J.~M.~Richard and P.~Sorba,
  Phys.\ Rev.\ D {\bf 73}, 054013 (2006)
  [hep-ph/0511039].

\bibitem{Buccella:2006fn}
  F.~Buccella, H.~H{\o}gaasen, J.~M.~Richard and P.~Sorba,
  Eur.\ Phys.\ J.\ C {\bf 49}, 743 (2007)
  [hep-ph/0608001].

\bibitem{Guo:2011gu}
  T.~Guo, L.~Cao, M.~Z.~Zhou and H.~Chen,
  arXiv:1106.2284 [hep-ph].


\bibitem{Hogaasen:2013nca}
  H.~H{\o}gaasen, E.~Kou, J.~M.~Richard and P.~Sorba,
  Phys.\ Lett.\ B {\bf 732}, 97 (2014)
  [arXiv:1309.2049 [hep-ph]].


\bibitem{Stancu:2009ka}
  F.~Stancu,
  J.\ Phys.\ G {\bf 37}, 075017 (2010)
  [arXiv:0906.2485 [hep-ph]].


\bibitem{Weinberg:1962hj}
  S.~Weinberg,
  Phys.\ Rev.\  {\bf 130}, 776 (1963).

\bibitem{Tornqvist:2004qy}
  N.~A.~T\"ornqvist,
  Phys.\ Lett.\ B {\bf 590}, 209 (2004)
  [hep-ph/0402237].

\bibitem{Voloshin:2003nt}
  M.~B.~Voloshin,
  Phys.\ Lett.\ B {\bf 579}, 316 (2004)
  [hep-ph/0309307].

\bibitem{Ding:2008gr}
  G.~J.~Ding,
  Phys.\ Rev.\ D {\bf 79}, 014001 (2009)
  [arXiv:0809.4818 [hep-ph]].

\bibitem{Wang:2013cya}
  Q.~Wang, C.~Hanhart and Q.~Zhao,
  Phys.\ Rev.\ Lett.\  {\bf 111},  132003 (2013)
  [arXiv:1303.6355 [hep-ph]].

\bibitem{Ma:2014zva}
  L.~Ma, X.~H.~Liu, X.~Liu and S.~L.~Zhu,
  Phys.\ Rev.\ D {\bf 91},  034032 (2015)
  [arXiv:1406.6879 [hep-ph]].


\bibitem{Wang:2013kra}
  Q.~Wang, M.~Cleven, F.-K.~Guo, C.~Hanhart, U.-G.~Mei\ss{}ner, X.~G.~Wu and
  Q.~Zhao,
  Phys.\ Rev.\ D {\bf 89},  034001 (2014)
  [arXiv:1309.4303 [hep-ph]].

\bibitem{Voloshin:2011qa}
  M.~B.~Voloshin,
  Phys.\ Rev.\ D {\bf 84}, 031502 (2011)
  [arXiv:1105.5829 [hep-ph]].

\bibitem{Cleven:2011gp}
  M.~Cleven, F.-K.~Guo, C.~Hanhart and U.-G.~Mei\ss ner,
  Eur.\ Phys.\ J.\ A {\bf 47}, 120 (2011)
  [arXiv:1107.0254 [hep-ph]].

\bibitem{Hanhart:2010wh}
  C.~Hanhart, Y.~S.~Kalashnikova and A.~V.~Nefediev,
  Phys.\ Rev.\ D {\bf 81}, 094028 (2010)
  [arXiv:1002.4097 [hep-ph]].

\bibitem{Filin:2010se}
  A.~A.~Filin, A.~Romanov, V.~Baru, C.~Hanhart, Y.~S.~Kalashnikova,
  A.~E.~Kudryavtsev, U.-G.~Mei\ss{}ner and A.~V.~Nefediev,
  Phys.\ Rev.\ Lett.\  {\bf 105}, 019101 (2010)
  [arXiv:1004.4789 [hep-ph]].

\bibitem{Guo:2011dd}
  F.-K.~Guo and U.-G.~Mei\ss{}ner,
  Phys.\ Rev.\ D {\bf 84}, 014013 (2011)
  [arXiv:1102.3536 [hep-ph]].

\bibitem{Nieves:2012tt}
  J.~Nieves and M.~P.~Valderrama,
  Phys.\ Rev.\ D {\bf 86}, 056004 (2012)
  [arXiv:1204.2790 [hep-ph]].

\bibitem{HidalgoDuque:2012pq}
  C.~Hidalgo-Duque, J.~Nieves and M.~P.~Valderrama,
  Phys.\ Rev.\ D {\bf 87}, 076006 (2013)
  [arXiv:1210.5431 [hep-ph]].

\bibitem{Guo:2013sya}
  F.-K.~Guo, C.~Hidalgo-Duque, J.~Nieves and M.~P.~Valderrama,
  Phys.\ Rev.\ D {\bf 88}, 054007 (2013)
  [arXiv:1303.6608 [hep-ph]].

\bibitem{Fleming:2007rp}
  S.~Fleming, M.~Kusunoki, T.~Mehen and U.~van Kolck,
  Phys.\ Rev.\ D {\bf 76}, 034006 (2007)
  [hep-ph/0703168].

\bibitem{Nieves:2011vw}
  J.~Nieves and M.~P.~Valderrama,
  Phys.\ Rev.\ D {\bf 84}, 056015 (2011)
  [arXiv:1106.0600 [hep-ph]].

\bibitem{Valderrama:2012jv}
  M.~P.~Valderrama,
  Phys.\ Rev.\ D {\bf 85}, 114037 (2012)
  [arXiv:1204.2400 [hep-ph]].

\bibitem{Baru:2015nea}
  V.~Baru, E.~Epelbaum, A.~A.~Filin, F.-K.~Guo, H.-W.~Hammer, C.~Hanhart,
  U.-G.~Mei\ss{}ner and A.~V.~Nefediev,
  Phys.\ Rev.\ D {\bf 91}, 034002 (2015)
  [arXiv:1501.02924 [hep-ph]].

\bibitem{Alhakami:2015uea}
  M.~H.~Alhakami and M.~C.~Birse,
  Phys.\ Rev.\ D {\bf 91}, 054019 (2015)
  [arXiv:1501.06750 [hep-ph]].

\bibitem{Braaten:2015tga}
  E.~Braaten,
  arXiv:1503.04791 [hep-ph].

\bibitem{Tornqvist:1993ng}
  N.~A.~T\"ornqvist,
  Z.\ Phys.\ C {\bf 61}, 525 (1994)
  [hep-ph/9310247].

\bibitem{Zhang:2006ix}
  Y.~J.~Zhang, H.~C.~Chiang, P.~N.~Shen and B.~S.~Zou,
  Phys.\ Rev.\ D {\bf 74}, 014013 (2006)
  [hep-ph/0604271].

\bibitem{Thomas:2008ja}
  C.~E.~Thomas and F.~E.~Close,
  Phys.\ Rev.\ D {\bf 78}, 034007 (2008)
  [arXiv:0805.3653 [hep-ph]].

\bibitem{Liu:2008tn}
  X.~Liu, Z.~G.~Luo, Y.~R.~Liu and S.~L.~Zhu,
  Eur.\ Phys.\ J.\ C {\bf 61}, 411 (2009)
  [arXiv:0808.0073 [hep-ph]].

\bibitem{Liu:2008mi}
  Y.~R.~Liu and Z.~Y.~Zhang,
  Phys.\ Rev.\ C {\bf 80}, 015208 (2009)
  [arXiv:0810.1598 [hep-ph]].

\bibitem{Molina:2009ct}
  R.~Molina and E.~Oset,
  Phys.\ Rev.\ D {\bf 80}, 114013 (2009)
  [arXiv:0907.3043 [hep-ph]].

\bibitem{Lee:2009hy}
  I.~W.~Lee, A.~Faessler, T.~Gutsche and V.~E.~Lyubovitskij,
  Phys.\ Rev.\ D {\bf 80}, 094005 (2009)
  [arXiv:0910.1009 [hep-ph]].

\bibitem{Sun:2011uh}
  Z.~F.~Sun, J.~He, X.~Liu, Z.~G.~Luo and S.~L.~Zhu,
  Phys.\ Rev.\ D {\bf 84}, 054002 (2011)
  [arXiv:1106.2968 [hep-ph]].

\bibitem{Ohkoda:2011vj}
  S.~Ohkoda, Y.~Yamaguchi, S.~Yasui, K.~Sudoh and A.~Hosaka,
  Phys.\ Rev.\ D {\bf 86}, 014004 (2012)
  [arXiv:1111.2921 [hep-ph]].

\bibitem{Ozpineci:2013zas}
  A.~Ozpineci, C.~W.~Xiao and E.~Oset,
  Phys.\ Rev.\ D {\bf 88}, 034018 (2013)
  [arXiv:1306.3154 [hep-ph]].

\bibitem{He:2013nwa}
  J.~He, X.~Liu, Z.~F.~Sun and S.~L.~Zhu,
  Eur.\ Phys.\ J.\ C {\bf 73}, 2635 (2013)
  [arXiv:1308.2999 [hep-ph]].

\bibitem{Aceti:2014kja}
  F.~Aceti, M.~Bayar, J.~M.~Dias and E.~Oset,
  Eur.\ Phys.\ J.\ A {\bf 50}, 103 (2014)
  [arXiv:1401.2076 [hep-ph]].

\bibitem{Aceti:2014uea}
  F.~Aceti, M.~Bayar, E.~Oset, A.~Martinez Torres, K.~P.~Khemchandani, J.~M.~Dias, F.~S.~Navarra and M.~Nielsen,
  Phys.\ Rev.\ D {\bf 90},  016003 (2014)
  [arXiv:1401.8216 [hep-ph]].

\bibitem{Zhao:2014gqa}
  L.~Zhao, L.~Ma and S.~L.~Zhu,
  Phys.\ Rev.\ D {\bf 89}, 094026 (2014)
  [arXiv:1403.4043 [hep-ph]].

\bibitem{Liu:2008xz}
  X.~Liu, Y.~R.~Liu, W.~Z.~Deng and S.~L.~Zhu,
  Phys.\ Rev.\ D {\bf 77}, 094015 (2008)
  [arXiv:0803.1295 [hep-ph]].

\bibitem{Ding:2008mp}
  G.~J.~Ding, W.~Huang, J.~F.~Liu and M.~L.~Yan,
  Phys.\ Rev.\ D {\bf 79}, 034026 (2009)
  [arXiv:0805.3822 [hep-ph]].

\bibitem{Li:2015exa}
  M.~T.~Li, W.~L.~Wang, Y.~B.~Dong and Z.~Y.~Zhang,
  Commun.\ Theor.\ Phys.\  {\bf 63}, 63 (2015).

\bibitem{grossbuch}
F.~Gross, {\it Relativistic Quantum Mechanics and Field Theory}, John Wiley \& Sons, New York, 1993.

\bibitem{Ericson:1993wy}
  T.~E.~O.~Ericson and G.~Karl,
  Phys.\ Lett.\ B {\bf 309}, 426 (1993).

\bibitem{Tornqvist:1991ks}
  N.~A.~T\"ornqvist,
  Phys.\ Rev.\ Lett.\  {\bf 67}, 556 (1991).


\bibitem{Albaladejo:2015dsa}
  M.~Albaladejo, F.-K.~Guo, C.~Hidalgo-Duque, J.~Nieves and M.~P.~Valderrama,
  arXiv:1504.00861 [hep-ph].

\bibitem{Hanhart:2014ssa}
  C.~Hanhart, J.~R.~Pel\'aez and G.~R\'ios,
  Phys.\ Lett.\ B {\bf 739}, 375 (2014)
  [arXiv:1407.7452 [hep-ph]].

\bibitem{Wang:2014wga}
  Q.~Wang,
  Phys.\ Rev.\ D {\bf 89},  114013 (2014)
  [arXiv:1403.2243 [hep-ph]].


\end{thebibliography}
\end{document}